\documentclass[%
reprint,
amsmath,amssymb,
 aps,
twocolumn
]{revtex4-1}
\usepackage[x11names]{xcolor}
\usepackage{graphicx}
\usepackage{dcolumn}
\usepackage{bm}
\usepackage{braket}
\usepackage[utf8]{inputenc}
\usepackage{booktabs}
\usepackage{multirow}
\usepackage{tabularx}
\usepackage{tikz}
\usepackage{float}
\usepackage{placeins}
\usepackage{pgfplots}
\usetikzlibrary{patterns}
\newcommand*{\hham}{\hat{\mathcal{H}}}
\newcommand{\angstrom}{\textup{\AA}}
\newcommand{\tablefootnoteone}{\textup{ligand abbreviations: Me$_3$tacn = 1,4,7-trimethly-1,4,7-triazacyclonane; bpy = bipyridine; bpea = N,N-bis(2-pyridylmethyl)ethylamine; ma = 3-oxy-2-methyl-
4$H$-pyran-4-onato-$O^3,O^4$}}
\usepackage{array}
\pgfplotsset{compat=1.17}
\begin{document}
\title{Comparative Density Functional Theory Study \\of Magnetic Exchange Coupling in Di-nuclear Transition Metal Complexes}
\author{Henry C. Fitzhugh$^{1*}$}
\author{James W. Furness$^1$}
\author{Mark R. Pederson$^2$}
\author{Juan E. Peralta$^3$}
\author{Jianwei Sun$^1$}
    \email{hfitzhug@tulane.edu, jsun@tulane.edu}
\affiliation{\mbox{$^1$Department of Physics and Engineering Physics, Tulane University, New Orleans, Louisiana 70118, USA} \\ $^2$Department of Physics, the University of Texas at El Paso, El Paso, Texas 79968, USA \\ $^3$Department of Physics and Science of Advanced Materials, Central Michigan University, Mount Pleasant, Michigan 48859, USA.}
\date{\today}

\begin{abstract}
Multi-center transition metal complexes (MCTMs) with magnetically interacting ions 
have been proposed as components for 
information processing devices and storage units.
For any practical application of MCTMs as magnetic units, it is crucial to characterize their magnetic behavior, and in particular the isotropic magnetic exchange coupling, $J$, between its magnetic centers.
Due to the large size of typical MCTMs, density functional theory (DFT) is the only practical electronic structure method for evaluating the $J$ coupling. 
Here we assess the accuracy of different density functional approximations for predicting the magnetic couplings of seven di-metal transition metal complexes, including five di-manganese, one di-copper, and one di-vanadium with known reliable experimental $J$ couplings spanning from ferromagnetic to strong antiferromagnetic.
The density functionals considered include global hybrid functionals which mix semilocal density functional approximations and exact exchange with a fixed admixing parameter, six local hybrid functionals where the admixing parameters are extended to be spatially dependent, the SCAN and r$^2$SCAN meta-generalized gradient approximations (GGAs), and two widely used GGAs. 
We found that global hybrids have a tendency to over-correct the error in magnetic coupling parameters from the Perdew-Burke-Ernzerhof (PBE) GGA, while the performance of local hybrid density functionals is scattered without a clear trend, suggesting that more efforts are needed for the extension from global to local hybrid density functionals for this particular property.
The SCAN and r$^2$SCAN meta-GGAs are found to perform as well or better than the global and local hybrids on most tested complexes. We further analyze the charge density redistribution of meta-GGAs as well as global and local hybrid density functionals with respect to that of PBE, in connection to the self-interaction error (SIE) or delocalization error.
\end{abstract}

\maketitle

\section{Introduction}
Multi-center transition metal complexes (MCTMs) can be used as single molecule magnets (SMMs), which
exhibit a purely molecular magnetic hysteresis and superparamagnetic behavior below a certain blocking temperature \cite{smm01, smm02}.
The electronic and nuclear spin-states of SMMs have been used for storage and manipulation of quantum information, constituting qubits in several candidate computational architectures \cite{smm01, smm02}.
A one-SMM device has been made to execute a simplest case of Grover's algorithm \cite{Ruben}.
Other applications for SMMs include high-density binary (classical) storage and spintronic technology \cite{Hao, bg15}.

Efficient methods for analyzing the magnetic properties of large MCTMs are necessary for the targeted development of SMM-based technologies.
Prediction of key quantities such as the parameters of spin-Hamiltonian representations of magnetic structure will potentially allow researchers to identify and design candidate quantum materials.
Methods for modeling MCTM SMMs must reliably capture magnetic interactions between magnetic centers and generate accurate magnetic coupling constants $J$ for use in a Heisenberg--Dirac--van Vleck model Hamiltonian, 
\begin{equation}
    \hham=-2\sum_{i>j}J_{ij}\mathbf{S}_i\cdot\mathbf{S}_j,
    \label{spinHgeneral}
\end{equation}
where the unpaired electrons for each metal ion center are approximated as a single spin center $\mathbf{S}_i$. Parameters $J_{ij}$ describe the strength and nature of the pairwise interactions; e.g. negative and positive $J$ value indicates anti-ferromagnetic and ferromagnetic interactions, respectively.

The most accurate extraction of a realistic magnetic Hamiltonian from electronic structure calculations  requires a suitably large configuration space and multi-reference techniques \cite{mrm1, mrm2, mrm3, mrm4, mrm5, mrm6, mrm7}.
Multi-reference methods scale poorly with the number of electrons, additionally, the practical need of a limited active space of reference states lead to somewhat ad-hoc variations in criteria for reference state inclusion.
Density functional theory (DFT) with the so-called \emph{broken symmetry} techniques \cite{Noodleman, RuizAlvarez, Yamaguchi, bs1, bs2, bs3, bs4, bs5, bs6, bs7, bs8, bs9} offers efficiency gains and potential for a reduced number of reference calculations.
Using DFT, an intractable multi-reference problem can be reduced to the manageable task of extracting effective exchange-coupling constants between interacting magnetic centers in a pairwise manner as embodied in a nearest-neighbors spin-Hamiltonian representation, allowing DFT to tackle large nuclearity complexes with ease. \cite{lee2022synthesis}.

\begin{figure*}
    \centering
    \fbox{\includegraphics[width=\textwidth]{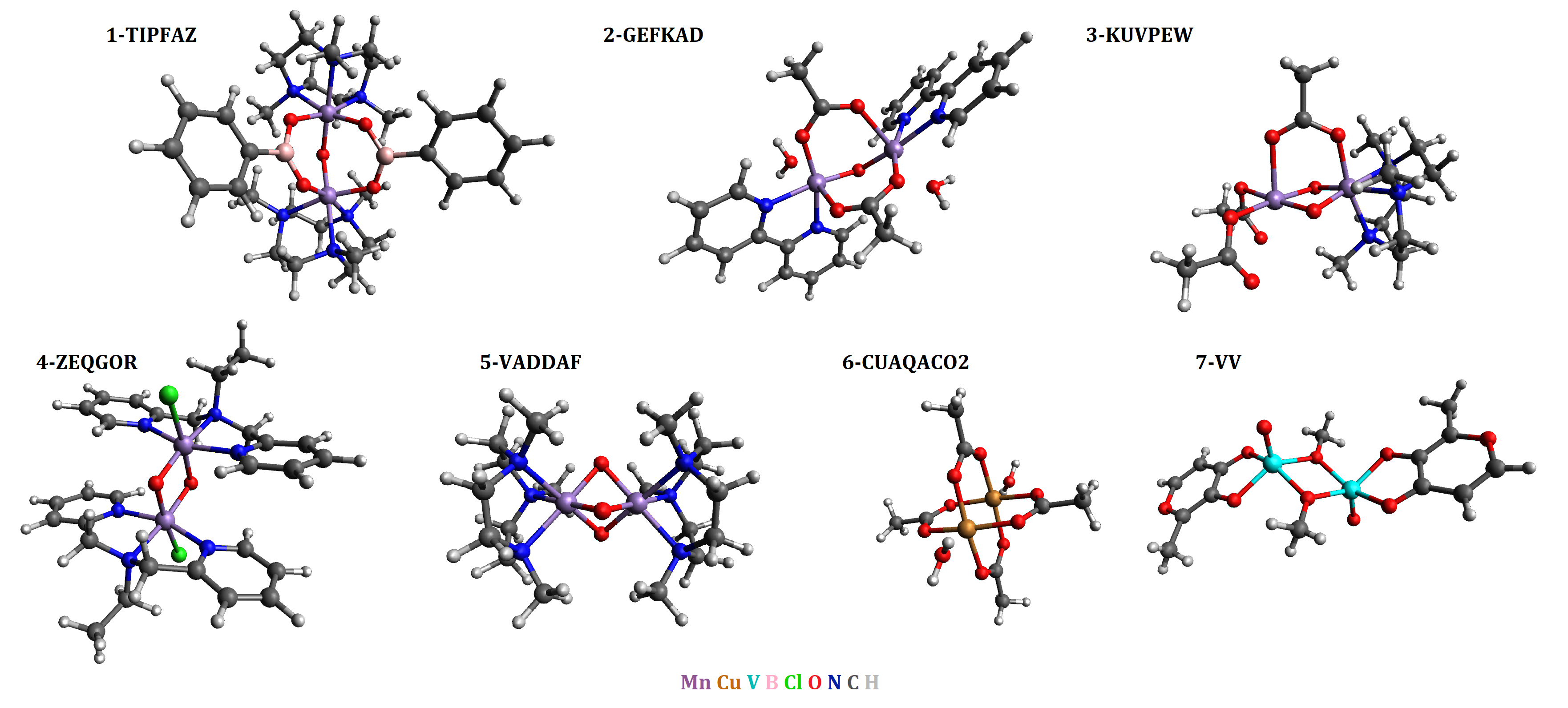}}
    \caption{Structures for all seven transition metal complexes in the study.}
    \label{fig:test_set}
\end{figure*}

\begin{table*}[t]
    \centering
    \caption{\label{tab:table1}
    Here we list and characterize the dinuclear transition metal complexes considered in this study. From left to right, the columns of the table provide: an assigned number, moniker which is the Cambridge Crystallographic Database reference code for complexes 1-5, structural formula, metal (M) oxidation states, formal spin at each M, M-M distance $R$ (in $\angstrom$), experimentally determined exchange constant $J^{exp.}$ (in $cm^{-1}$), and references for experimental $J^{exp}$. Each complex will be referenced by its respective assigned number and moniker.}
    \begin{ruledtabular}
    \begin{tabular}{ccccccccr}
    \textrm{number}&\textrm{moniker}&\textrm{Structural Formula\footnote{\tablefootnoteone}}&\textrm{Oxi. States}&\textrm{$\mathrm{S}_1$}&\textrm{$\mathrm{S}_2$}&\textrm{$R$}&\textrm{$J^{exp.}$}&\textrm{Ref.}\\
    \colrule
    1 & TIPFAZ & $\mathrm{\small{[Mn_2O(O_2BPh)_2(Me_3tacn)_2](PF_6)_2}}$ 
     & IV, IV & $3/2$ & $3/2$ & 3.185 & 10 & \cite{MN1}\\
    2 & GEFKAD & $\mathrm{\small{[Mn_2O(OAc)_2(H_2O)_2(bpy)_2](PF_6)_2}, 1.75H_2O}$
     & III, III & $2$ & $2$ & 3.131 & -3.4 & \cite{MN2}\\
    3 & KUVPEW & $\mathrm{\small{[Mn_2O_2(OAc)(Me_3tacn)(OAc)_2]}}$
     & III, IV & $2$ & $3/2$ & 2.665 & -90 & \cite{MN3}\\
    4 & ZEQGOR & $\mathrm{\small{[Mn_2O_2Cl_2(bpea)_2](ClO_4)_2}}$
     & IV, IV & $3/2$ & $3/2$ & 2.756 & -147 & \cite{MN4}\\
    5 & VADDAF & $\mathrm{\small{[Mn_2O_3(Me_3tacn)_2](PF_6)_2, H_2O}}$
     & IV, IV & $3/2$ & $3/2$ & 2.297 & -390 & \cite{MN5}\\
    \colrule
    6 & CUAQACO2 & $\mathrm{\small{[\{Cu(H_2O)\}_2(\mu -AcO)_4]}}$
     & II, II & $1/2$ & $1/2$ & 2.616 & -286 & \cite{Cu2}\\
    7 & VV & $\mathrm{\small{[(\mu-OCH_3)VO(ma)]_2}}$
     & IV, IV & $1/2$ & $1/2$ & 3.081 & -107 & \cite{V2}\\
    \end{tabular}
    \end{ruledtabular}
\end{table*}

The overall accuracy of any DFT calculation is largely determined by the choice of the exchange-correlation (XC) functional. 
This is especially apparent for calculations of magnetic properties, in part because the differences in energy between two magnetic states is often small. 
In practice, the value and even the sign of $J$ can vary widely with the choice of the XC functional, potentially leading to qualitatively different results.
Given this extreme sensitivity to XC functional choice, this paper aims to analyze the performance of functionals from across the Perdew--Schmidt hierarchy \cite{JacobsLadder} for a set of seven dinuclear transition metal organic complexes.

Numerous comparisons of density functionals for magnetic dimers have been previously performed \cite{Comba2009calculation, bs6, OrioHybrid, joshi2020accuracy, bovi2012magnetic, Cirera, PhillipsPeraltaMolSet, costa2018post, Pantazis, schinzel2009validation}.
This work intends to expand the analysis to include local hybrid functionals and the SCAN family of functionals, both of which have shown promise for magnetic systems \cite{LHFrecentreview, doi:10.1063/1.3596070, janesko2021replacing, holzer2021assessing}\cite{PhysRevB.107.045126, fu2019density, TrickeySPinCrossover, pokharel2022sensitivity, PhysRevB.105.195134, mejia2019analysis, chakraborty2018predicting}.

\section{The Seven Metal-Organic Complexes}
The seven dimers of the test set are shown in Figure \ref{fig:test_set}.
The experimental couplings, $J$, were determined by magnetic susceptibility measurements \cite{MN1, MN2, MN3, MN4, MN5, Cu2, V2}.
As other authors have noted \cite{PhillipsPeraltaMolSet}, the development of a large and inclusive test set that features a variety of 3d metals could be beneficial for the understanding and improvement of functionals for use in MCTM research.
The test set of the present analysis constitutes a small sample from the overall space of MCTM configurations, but it is representative of a number of relevant cases \cite{Pantazis}.

The five manganese dimers vary from the weakly ferromagnetic 1-TIPFAZ ($J = 10$ $cm^{-1}$) to the strongly anti-ferromagnetic 5-VADDAF ($J = -390$ $cm^{-1}$), see table \ref{tab:table1}.
The manganese dimers were used by D. Pantazis for a similar performance assessment of double hybrid functionals \cite{Pantazis}, and have all been studied extensively for magnetic properties.
1-TIPFAZ, 4-ZEQGOR, and 5-VADDAF are Mn(IV)--Mn(IV) complexes.
2-GEFKAD has an oxidation state of Mn(III)--Mn(III) while 3-KUVPEW has mixed valence with oxidation of Mn(III)--Mn(IV).
The 3-KUVPEW model dimer has neutral overall charge, while the other four dimers have $+2$ overall charge.
Our test set has origins in biochemistry research where manganese dimers are important bio-mimetic analogs of protein complexes. Manganese complexes are important for numerous biochemical processes such as the oxygen-evolving complex of photosystem II \cite{bg1, bg2, bg3, bg4, bg5, bg6, bg7, bg8, bg9, bg10, bg11, bg12, bg13, bg14}.

The paddle-wheel di-copper complex 6-CUAQACO2 and di-vanadium complex 7-VV provide contrast to the manganese results.
While the manganese atoms in the five complexes all feature multiple unpaired electrons per spin center where oxidation is either IV ($S=3/2$) or III ($S=2$), the two remaining complexes each have a single unpaired electron per metal ion; 6-CUAQACO2 copper ions have a nearly full d-shell with one unpaired electron, and 7-VV has a single d-shell electron per vanadium ion.

\section{The Functionals of the Study}
Density functional approximations can be broadly categorized into the Perdew-Schmidt hierarchy of increasing sophistication, often called the ``Jacob's ladder'' of density functional theory \cite{JacobsLadder}. 
The lowest three rungs of the ladder contain XC functionals depending only on (semi-) local ingredients: local density approximations (LDA) including only the electron density, generalized-gradient approximations (GGA) that also include the electron density gradient, and the meta-GGA, which expands this to include additional semi-local ingredients such as the kinetic energy density, $\tau(\mathbf{r})$, or the density Laplacian.
Higher rungs incorporate further complexity through direct non-local dependence on the occupied Kohn--Sham orbitals as single Slater-determinant (SSD) exchange (frequently called ``exact'' or ``Hartree--Fock'' exchange) as in hybrid functionals.
Beyond hybrid functionals, higher rungs may introduce additional dependence on the unoccupied Kohn--Sham orbitals, although the associated computational cost prohibits their use for MCTM studies and hence they will not be considered here.
In general one can expect accuracy to improve as one climbs to  higher rungs, but at the expense of increasing computational cost. 

Reduction of self-interaction (SIE) is a key reason for improvement of functional performance as one ascends to higher rungs in the hierarchy \cite{LundbergSIEeffects, RuizAlvarez}.
When SIE is present, the spurious self-Coulomb interaction does not cancel out with the spurious self-exchange-correlation interaction, and is responsible for the slight delocalization of the un-paired electrons. 
In the systems discussed here, the unpaired electrons at the metal sites have 3d character. 
While the spurious delocalization introduced by SIE may be viewed as small, the overlap between the d-electrons and bridging atoms is due to the decaying atomic orbital and therefore the $J$ parameter scales exponentially with this overlap \cite{PhysRevB.68.020405}. 
Similarly, the self-interaction-induced delocalization of unpaired electrons in simple stretched-bond molecules leads to errors in transition from closed-shell to open-shell behavior (Coulson-Fischer point \cite{coulson1949xxxiv}) and the exchange-coupling parameter. 
The possibility of improving such coupling within the self-interaction-corrected methods was noted early in simple applications to Li$_2$ \cite{Li_2paper} and has been discussed in Refs. \cite{DYK2017,jonsson2021}. 
In Ref. \cite{DYK2017}, an application of the Fermi-L\"owdin orbital self-interaction corrected (FLOSIC) \cite{PRP2014} method identified both the relative energetic ordering and changes in localization for different d-orbital fillings in undercoordinated spin-carrying transition porphyrin.  
Very recently Hooshmand {\em et al} have applied FLOSIC to understand open-shell behavior in ozone which has similar characteristics as Refs \cite{Li_2paper,jonsson2021}. 
This work identified a case where even with the FLOSIC the assumptions of near-orthogonality between degenerate $\uparrow\downarrow$ and $\downarrow\uparrow$ states may require additional corrections that are ordinarily not included in broken symmetry methods. 

The accuracy in predicting $J$ can be connected with SIE by checking an approximate equation for magnetic coupling $J$ \cite{calzado2002analysis, calzado2002analysis2, cabrero2002metal},
\begin{equation} \label{J-contributions}
    J = J_\mathrm{FM} + J_\mathrm{AFM} = 2K - \frac{4t^2}{U}
\end{equation}
where $2K$ comprises the ferromagnetic contribution, $J_\mathrm{FM}$, of direct exchange between the orbitals of each spin-center, $t$ is the electron-transfer integral between magnetic orbitals, and $U$ is the on-site repulsive interaction of two electrons in a particular magnetic orbital.
It has been shown that for the prediction of $J$ for anti-ferromagnetic dimers featuring well-localized magnetic orbitals, $U$ is more sensitive than $t$ or $K$ to the choice of electronic structure methods \cite{calzado2002analysis, calzado2002analysis2, cabrero2002metal, PhillipsPeraltaMolSet}.
While semi-local functionals have SIE causing delocalization of electrons around the magnetic centers and thus an underestimation of $U$, the admixture of exact exchange tends to overlocalize the electrons, dramatically increasing $U$ and thus significantly decreasing the $J_\mathrm{AFM}$ contribution to $J$.

Although self-interaction correction \cite{PerdewZunger} can greatly correct delocalization errors, the size of MCTMs can hinder use due to how the method scales with the number of treated orbitals, with the newer FLOSIC showing promising improvements \cite{PRP2014}.
Thus, it is important to understand the accuracy and overall tendencies of XC functionals prior to any error-correction or additional calculations.

To examine accuracy of density functionals for $J$ and the effects of SIE, we have selected representative GGA, meta-GGA, and hybrid functionals, chosen as widely used examples of their respective classes.
For semi-local functionals, some degree of SIE is unavoidable as the semi-local exchange is unable to exactly recover the non-local behavior required. In practice, the degree of SIE varies between functionals, and sophisticated meta-GGAs have typically shown a marked reduction of SIE compared to simpler GGAs \cite{10.1103/PhysRevB.102.045112, 10.1063/1.5055623, PhillipsPeraltaMolSet}.

The PBE and BLYP functionals were chosen as representative GGAs, with PBE serving as a baseline of comparison of results. The SIE of PBE and resulting delocalization and error in $J$ are more severe than those of all higher-rung functionals considered. 

We take the SCAN (Strongly Constrained and Approriately Normed) \cite{scan}, the r$^2$SCAN \cite{doi:10.1021/acs.jpclett.0c02405}, and the TPSS \cite{TPSS} functionals as examples from the meta-GGA rung.
The SCAN functional broadly builds upon the earlier TPSS functional and was created to satisfy all additional mathematical constraints that the exact XC is known to obey.
SCAN has shown high general accuracy for a wide range of properties and systems \cite{59205668, 75135662, SCANhightempSC, 362159062, PhysRevB.98.125140, 382089150, 155821068}.
Despite this general success, extensive use has revealed SCAN to be numerically problematic in many situations, and the
r$^2$SCAN functional was developed to eliminate these numerical difficulties \cite{doi:10.1021/acs.jpclett.0c02405}.
r$^2$SCAN maintains much of SCAN's good general accuracy under testing \cite{ehlert2021r2scan, grimme2021r2scan}, even showing mild improvement over SCAN in some domains.
This improved numerical performance comes at the cost of the an incorrect fourth-order term in the slowly-varying density gradient expansion of r$^2$SCAN, which is recovered exactly by the original SCAN functional, such that r$^2$SCAN obeys one fewer exact constraint than SCAN \cite{10.1063/5.0073623}.

A recent study of functional performance for spin-crossover systems, where results similarly depend on small energy differences between magnetic states, suggests that r$^2$SCAN can accurately predict magnetic spin-crossover properties \cite{TrickeySPinCrossover} further motivating its inclusion in the present study.
De-orbitalized versions of SCAN and r$^2$SCAN, SCAN-L and r$^2$SCAN-L \cite{mejia2017deorbitalization, mejia2018deorbitalized, tran2018orbital}, performed well in this spin-crossover study.
It will be interesting to observe their performance for magnetic molecular systems, but they are not yet included in Turbomole which was used for the present study. 
Some meta-GGA functionals behave similarly to GGA functionals for the calculation of magnetic properties of MCTM systems, leading to dismissal of functionals lacking an exact exchange component \cite{OrioHybrid, OrioTPSSh, Cirera}.
Given the success of SCAN and r$^2$SCAN for other magnetic properties, this broad assessment of meta-GGAs for broken symmetry calculations of magnetic coupling parameters will be reassessed.

Beyond the semi-local level, we examine the performance of global hybrids B3LYP \cite{b3lyp} and TPSSh \cite{tpssh}, as the two most widely used functionals in MCTM studies.
A \emph{global} hybrid XC functional includes a proportion of SSD exchange alongside a scaled semi-local exchange functional, with a constant mixing parameter $a_0$ controlling the mixing ratio.
In the simplest arrangement for a global hybrid,
\begin{equation}
    E^\mathrm{GH}_\mathrm{XC} = a_0E_\mathrm{X}^\mathrm{SSD}+(1 - a_0)E^\mathrm{DFT}_\mathrm{X}+E_\mathrm{C}. \label{eq:global_hyb}
\end{equation}
In previous comparisons of functional performance for manganese dimers, TPSSh produced the most accurate DFT-based broken symmetry calculations of $J$ \cite{OrioTPSSh, OrioHybrid, Pantazis} and will here be considered a benchmark for comparison to meta-GGAs and local hybrid functionals.
While often highly accurate for specialized applications, global hybrid functionals have proven to be generally inflexible for universal accuracy.
A fixed proportion of SSD exchange suitable for one empirical domain often significantly lessens the accuracy in another.
While $10-25\%$ SSD exchange appears to provide the best thermochemical accuracy in many cases, such functionals show a tendency to underestimate reaction barriers \cite{PBE0, doi:10.1021/jp049908s}.
Hydrogen transfer reactions, for example, are improved significantly by larger fractions of SSD exchange than would allow accurate thermochemistry \cite{hydrogenhybrid}.
A larger SSD-exchange admixture tends to improve predictions of electron paramagnetic resonance (EPR) parameters for MCTMs but deteriorates performance for main group radicals \cite{kaupp2002calculation}.

When calculating magnetic couplings with the broken symmetry method, researchers have found that GGA and some meta-GGA functionals tend to over-stabilize low-spin ground states, whereas their hybrid counterparts over-stabilize the higher-multiplicity excited states \cite{OrioTPSSh}. 
The hybrid meta-GGA TPSSh is thought to achieve a balance between these competing factors for many transition metal groups, and is hence a recommended XC functional for studying MCTMs and related systems \cite{OrioTPSSh, OrioHybrid, Cirera}.  
Despite the successes of TPSSh, some transition metal magnetic systems benefit from a higher proportion of SSD exchange.
For example, when calculating $J$ for chromium dimers that are isoelectric to manganese dimers, a significantly larger admixture of SSD exchange was surprisingly needed to produce accurate values \cite{pantazis2019meeting}.
Fe(III) complexes provide another example where a larger (25\%) admixture of SSD exchange improves the reliability of calculated J couplings \cite{joshi2020accuracy}.
Thus, calculations of magnetic coupling parameters using global hybrids has tended to rely on a \textit{just enough but not too much} ``Goldilocks'' style of reasoning in which SSD exchange admixture is determined ad hoc depending on the elements and groups present \cite{PhillipsPeraltaMolSet, cramer2009density} and TPSSh emerges as best-case for a generic ``just right'' functional; though there is little theoretical basis for such choices and the reasons for exceptions to ``Goldilocks'' generalities don't have clear or reliably consistent explanations.

Local hybrid functionals can be considered a generalization of Eq. \ref{eq:global_hyb} that permit the mixing parameter $a_0$ to vary as a function of space.
By controlling the mix of DFT and SSD exchange as a local function, it is expected that SSD exchange can be targeted to correct SIE where it is present, while avoiding inaccuracies caused by the mismatched locality of the exchange and correlation holes \cite{Jaramillo2003}. In principle, this could eliminate the need for ad hoc admixtures in order to improve accuracy.
\begin{equation}
    E^\mathrm{LH}_\mathrm{XC} = \int{a(\mathbf{r})\epsilon_\mathrm{X}^\mathrm{SSD}(\mathbf{r})d\mathbf{r}}+\int{[1-a(\mathbf{r})][\epsilon^\mathrm{DFT}_\mathrm{X}(\mathbf{r})]d\mathbf{r}} + E_\mathrm{C} \label{eq:local_hyb}
\end{equation}
The function $a(\mathbf{r})$ that governs the mix of SSD and semi-local exchange typically, though not necessarily, spans the range $[0,1]$. Reviews of local hybrid functionals and reasons for their development can be found in several articles \cite{LHFrecentreview, janesko2021replacing, 20t}.

Currently, local hybrid functionals have not seen wide adoption; however their potential to control SIE in an affordable first principles DFT model makes them of special interest for MCTM studies. Consequently, we consider six local hybrids here.

One of the first local hybrids proposed has a local mixing function $a(\mathbf{r})=z(\mathbf{r}) = \tau_W(\mathbf{r})/\tau(\mathbf{r})$, a quantity used in the TPSS functional and other meta-GGAs, where $\tau_W(\mathbf{r})=|\nabla\rho|^2/8\rho$ is the von Weiz\"acker kinetic-energy density \cite{Jaramillo2003} and $\tau(\mathbf{r})=\frac{1}{2}\Sigma^{occ.}_i|\nabla\phi_i(\mathbf{r})|^2$ is the KS orbital kinetic energy density.
$\tau_W$ is exactly equal to $\tau$ in one-electron regions and approaches zero in regions of slowly varying densities. 
The resulting mixing function selects full semi-local exchange when the local band gap is metallic in character and full SSD exchange in single-orbital regions where SIE can be exactly removed.
In the local hybrid lh07t-svwn \cite{07t}, the quantity $z(\mathbf{r})$ from Jaramillo et al. is scaled by a parameter optimized by the evaluation of atomization energies of the G2-1 set \cite{G1, G2}.
The resulting semi-empirical mixing parameter, $a(\mathbf{r})=b\:z(\mathbf{r})$ with $b=0.48$, no longer completely eliminates single-orbital SIE, but performs better on the G2-1 test set than global hybrids with typical optimized global mixing parameter $a_0$ near 0.2-0.25 \cite{07t}.

The local hybrid lh07s-svwn \cite{07s}, replaces $z(\mathbf{r})$ with a GGA mixing function that uses the dimensionless gradient of electron density, 
\begin{equation}
    s(\mathbf{r}) = |\nabla\rho| / 2k_F\rho \equiv |\nabla \rho| / 2(3\pi^2)^{1/3}\rho^{4/3}.
\end{equation}
To construct a local mixing function $a(\mathbf{r})$, $s(\mathbf{r})$ is mapped to the interval $[0,1]$, where,
\begin{equation}
    a(\mathbf{r}) = \{s(\mathbf{r})/(\lambda + s(\mathbf{r}))\}^2
\end{equation} 
was chosen over other simple alternatives and parameterized against G2-1 atomization energies leading to a choice of $\lambda = 0.73$. 

The local hybrid lh12ct-ssirpw92 \cite{12ct} was constructed with a novel ansatz for the correlation portion of the functional that is based on partial elimination of the one-electron self-correlation for a short-range portion.
The range separation serves as a method for separating dynamic and non-dynamic correlation.
In lh12ct-ssi$r$pw92, a partial reduction in short range LSDA correlation is thought to help reduce SIE while maintaining empirical accuracy with calibration.
In lh12ct-ssi$f$pw92 the one-electron self-correlation is fully removed in the short range contribution \cite{12ct}.

While such local mixing initially appears simple, it introduces a complication through the semi-local and SSD exchange energy densities lacking a common gauge.
This local variation over the space of integration produce a spurious gauge term that does not integrate to zero.
These product terms within the XC integration are unique to local hybrids.

The local hybrids lh14t-calPBE \cite{14t} and lh20t \cite{20t} were constructed with calibration functions to bring SSD exact exchange and semi-local DFT exchange into a common gauge.
By altering only $\epsilon^\mathrm{DFT}_\mathrm{X}(\mathbf{r})$ in equation \ref{eq:local_hyb} to include an added calibration function, $G(\mathbf{r})$ such that $\int{G(\mathbf{r})d\mathbf{r} = 0}$, we can see that the integrand for $E^\mathrm{LH-cal}_\mathrm{XC}$ contains a product of the calibration function and the mixing function that does not trivially vanish to zero by construction as with global hybrids,
\begin{equation}
E^\mathrm{LH-cal}_\mathrm{X} = \int{\epsilon^\mathrm{LH}_\mathrm{X}(\mathbf{r})d\mathbf{r}} + \int{[1 -  a(\mathbf{r})]G(\mathbf{r})d\mathbf{r}},
\end{equation}
and
\begin{equation}
E^\mathrm{LH-cal}_\mathrm{X} = E^\mathrm{LH}_X - \int{a(\mathbf{r})G(\mathbf{r})d\mathbf{r}}.
\end{equation}
Consequently, the interplay of gauge and mixing functions can lead to wide variation in energy and electron density and can serve as a means for multi-parameter calibration.
Using only semi-local ingredients, Arbuznikov and Kaupp created a general \emph{partial integration gauge} (\emph{pig}) scheme for the generation of fully semi-local calibration terms that do not require additional calculations of quantities derived from SSD exchange.
We examine two functionals produced by this partial integration scheme; lh14t-calPBE \cite{14t} and lh20t \cite{20t}.

lh14t-calPBE with partial-integration-gauge `pig1' calibration function uses integration by parts once to create first-order correction terms \cite{14t}. The resulting set of parameters were optimized for thermochemical kinetics and measures of non-dynamical correlation.
This functional improved upon the reaction barrier inaccuracies of previous local hybrids while maintaining or improving other figures of merit \cite{14t}. 

lh20t \cite{20t} uses a revised PBE exchange and B95 meta-GGA correlation within the general `pig2' second-order calibration scheme, where partial integration is used twice.
The many resulting product terms and parameters offer greater flexibility than with lh14t-calPBE.
lh20t is shown to improve thermochemical data over standard GGA functionals while maintaining accuracy over a wide range of measures including main group energetics, electron delocalization of mixed-valence systems, and TDDFT excitations \cite{20t}.

\section{Computational Methods}
The structures of all seven molecules were taken from crystallographic studies without further relaxation.
Use of crystallographic coordinates ensures that the molecular geometry corresponds to the structures of the experimentally determined values for $J$ via magnetic susceptibility measurements, allowing experimental values for $J$ to be suitable for an assessment of the accuracy of DFT methods.

The manganese structures were provided by D. Pantazis who has studied the magnetic properties of these five molecules extensively \cite{Pantazis}.
Pantazis began with crystallographic coordinates as can be found in the Cambridge Structural Database \cite{Cambridge, MN1, MN2, MN3, MN4, MN5}.
He removed counter-ions and solvent molecules, added hydrogen atoms to complete proper coordination when necessary, and optimized the positions of hydrogen atoms with ORCA \cite{ORCA} using the TPSS functional \cite{TPSS} and D3 dispersion corrections \cite{Grimme} with all unpaired electrons spin-aligned.
For non-manganese complexes, solvent molecules are included only where necessary for the correct coordination of the metal ion \cite{V2, Cu2, Rudra, PhillipsPeraltaMolSet}.
No other nuclear positions from the crystalographic coordinates were altered or optimized.

All reported single-point DFT calculations were performed using the TURBOMOLE (v. 7.4) quantum chemistry program package \cite{Turbomole} with the exception of the functional lh20t for which the 7.6 development version of TURBOMOLE was used.
The def2-TZVP basis set and corresponding auxiliary basis set for the resolution of the identity approximation were used for all calculations \cite{basisset_f, basisset_j, auxset_l}.
The largest conventional grid setting of $7$ was used in all calculations so that all functionals were well-behaved and effects from variation of the grid were minimized.

The Heisenberg--Dirac--van Vleck Hamiltonian for isotropic bilinear coupling of two spin centers is assumed for all seven dimers of the study,
\begin{equation}
\hham = -2J\mathbf{S}_1\mathbf{S}_2
\end{equation}
where $\mathbf{S_1}$ and $\mathbf{S_2}$ are the total spin operators for the two metallic spin centers and $J$ is the magnetic coupling constant.
As in previous studies of magnetic transition metal complex dimers, including the five Mn-Mn molecules of the present study \cite{Pantazis}, Yamaguchi's equation, originally derived for $\mu$-oxo bridged dimers, is used to calculate $J$ \cite{Yamaguchi},

\begin{equation}
    J = -\frac{E_{\mathrm{HS}} - E_{\mathrm{BS}}}{\braket{S^2}_{\mathrm{HS}} - \braket{S^2}_{\mathrm{BS}}}.
\end{equation}
Here $E_{\mathrm{HS}}$ and $\braket{S^2}_{\mathrm{HS}}$ are the energy and total spin expectation value of the high spin state, and $E_{\mathrm{BS}}$ and $\braket{S^2}_{\mathrm{BS}}$ are the energy and total spin expectation value of the broken symmetry state.

This \emph{broken symmetry} technique affords a commonly used way to calculate $J$ from a manageable number of DFT calculations \cite{bs1, bs2, bs3, bs4, bs5, bs6, bs7, bs8, bs9, PhysRevB.69.014416}.
Only two states are required for calculation of the exchange-coupling parameter for our two-center transition metal complexes.
An unrestricted \textit{high spin} state with ferromagnetic spins for all unpaired electrons on both metal ions is initially converged.
Single-reference Slater determinant based methods such as KS-DFT cannot easily access the anti-ferromagnetic triplet states.
Despite this limitation, Noodleman and others have shown analytically for the case of dimers, energies and spin information from the \emph{broken symmetry} state (resembling and Ising-like configuration), where the unpaired spins of the orbitals at one of the two centers are flipped to the other co-linear spin channel, can serve as a meaningful basis for calculation of $J$ \cite{Noodleman, Yamaguchi, Dai1, Dai2}.

All converged high-spin and broken-symmetry states were verified by inspection of transition metal spin populations using natural population analysis (NPA) \cite{NPA} as implemented in TURBOMOLE.
NPA was also utilized to examine the variation in performance of the examined functionals for differences in relative charge localization, and relative differences in ionization. NPA results can be found in the supplementary information.

Convergence to the correct broken-symmetry states was facilitated by an initial enlargement of the HOMO-LUMO gap via an orbital shift setting and by a damping setting for SCF iterations (see the supplementary information for detailed settings).
The enlargement of the HOMO-LUMO gap prevented convergence to densities corresponding to spurious excited states, maintaining the correct occupation of orbitals by unpaired electrons.
Due to the sensitivity of $J$ to energies of the two states, convergence thresholds for total energy of at least $10^{-7}$ Hartree ($0.022$ $cm^{-1}$) were used for all calculations.
Additional SCF calculations with non-adjusted HOMO-LUMO gap and lessened SCF damping verified that complete relaxation into the target broken symmetry state was achieved.

\begin{table*}[t]
    \centering
    \caption{\label{tab:table2} Exchange coupling constants $J$$($cm$^{-1}$$)$ calculated with selected density functionals for the seven complexes are provided for comparison with experimental values (exp.). Mean signed error (MSE), mean absolute error (MAE), root-mean-square-error (RMSE), and mean absolute relative error (MARE) are provided for the five manganese compounds and again for all seven molecules. Best performers for MAE, RMSE, and MARE are printed with bold text and ranked for smallest error with an \emph{Olympic medal} color scheme; \textbf{\textit{\textcolor{Gold4}{italicized gold for 1st}}}, \textbf{\textcolor{LightCyan4}{silver for 2nd}}, and \textbf{\textcolor{Chocolate4}{bronze for 3rd}}.} 
    \begin{ruledtabular}
    \begin{tabular}{l|rrrrrrr|rr|rr|rr|rr}
    \textbf{Method}&\multicolumn{7}{c}{\textbf{Transition Metal Complex, $J$$($cm$^{-1}$$)$}}&\multicolumn{2}{c}{\textbf{MSE}}&\multicolumn{2}{c}{\textbf{MAE}}&\multicolumn{2}{c}{\textbf{RMSE}}&\multicolumn{2}{c}{\textbf{MARE}}\\
    \textrm{ }&\textrm{1}&\textrm{2}&\textrm{3}&\textrm{4}&\textrm{5}&\textrm{6}&\textrm{7}&\textrm{Mn$_2$}&\textrm{all 7}&\textrm{Mn$_2$}&\textrm{all 7}&\textrm{Mn$_2$}&\textrm{all 7}&\textrm{Mn$_2$}&\textrm{all 7}\\
    \colrule
    \emph{exp.} \cite{MN1, MN2, MN3, MN4, MN5, Cu2, V2}     & \emph{10}    & \emph{-3.4}  & \emph{-90} & \emph{-147} & \emph{-390}  & \emph{-286} & \emph{-107} & - & - & - & - & - & - & - & - \\
    \colrule
    BLYP             & -25.6 & -72.2 & -183 & -263 & -611  & -551 & -267 & -107.0 & -137.2 & 107.0 & 137.2 & 204.5 & 256.3 & 5.24 & 4.09 \\
    PBE              & -18.6 & -59.3 & -167 & -252 & -596 & -556 & -272  & -94.5 & -129.7 & 94.5 & 129.7 & 193.0 & 248.7 & 4.28 & 3.41\\
    \colrule
    TPSS             & -12.4 & -48.9 & -151 & -218 & -544 & -496 & -206 & -70.6 & -94.6 & 70.6 & 94.6 & 139.8 & 180.5 & 3.44 & 2.69 \\
    SCAN             & -4.0  & -19.7 & -110 & -148 & -381 & -298 & -116 & -8.4 & -8.9  & \textbf{\textcolor{Chocolate4}{12.1}} & \textbf{\textit{\textcolor{Gold4}{11.5}}} & \textbf{\textit{\textcolor{Gold4}{7.8}}} & \textbf{\textit{\textcolor{Gold4}{12.1}}} & 1.29 & 0.94   \\
    r$^2$SCAN        & 20.0  & -6.6  & -98 & -142 & -399  & -385 & -136 & -1.3 & -18.0 & \textbf{\textit{\textcolor{Gold4}{7.1}}} & 22.2 & 20.1 & 49.6 & \textbf{\textit{\textcolor{Gold4}{0.42}}} & \textbf{\textit{\textcolor{Gold4}{0.38}}}  \\
    \colrule
    B3-LYP           & 27.4  & -10.6 & -80 & -117 & -358  & -230 & -104 & 16.5 & 20.2 & 19.3 & 22.3 & 35.0 & 42.3 & \textbf{\textcolor{Chocolate4}{0.85}} & \textbf{\textcolor{Chocolate4}{0.64}}  \\
    PBE0             & 41.5  & 2.5   & -60 & -91  & -326  & -186 & -88  & 37.4 & 43.7 & 37.4 & 43.7 & 68.2 & 82.7 & 1.15 & 0.90 \\
    TPSSh            & 14.5  & -18.8 & -98 & -142 & -411  & -314 & -127 & -6.9 & -11.9 & \textbf{\textcolor{LightCyan4}{10.6}} & \textbf{\textcolor{LightCyan4}{14.5}} & \textbf{\textcolor{LightCyan4}{16.9}} & 24.1 & 1.03 & 0.77  \\
    \colrule
    lh07t-SVWN & 45.0  & 6.2   & -79 & -118 & -375  & -311 & -99  & 20.0 & 11.9 & 20.0 & 19.0 & 24.3 & \textbf{\textcolor{Chocolate4}{22.5}} & 1.33 & 0.98 \\
    lh07s-SVWN            & 42.4  & 3.4   & -86 & -126 & -376  & -318 & -109 & 15.6 & 6.5 & 15.6 & \textbf{\textcolor{Chocolate4}{15.9}} & \textbf{\textcolor{Chocolate4}{19.1}} & \textbf{\textcolor{LightCyan4}{18.8}} & 1.09 & 0.80 \\
    lh12ct-ssir         & 41.3  & -1.5  & -73 & -109 & -352  & -265 & -90  & 25.3 & 23.5 & 25.3 & 23.5 & 40.3 & 41.0 & \textbf{\textcolor{Chocolate4}{0.85}} & \textbf{\textcolor{Chocolate4}{0.64}} \\
    lh12ct-ssif         & 44.9  & 1.2   & -67 & -102 & -341  & -246 & -85  & 31.3 & 31.2 & 31.3 & 31.2 & 51.2 & 54.8 & 1.11 & 0.84 \\
    lh14t-calPBE  & 38.1  & 2.1   & -79 & -112 & -370  & -284 & -96  & 19.9 & 16.1 & 19.9 & 16.1 & 29.0 & 27.5 & 0.97 & 0.71 \\
    lh20t            & 35.9  & -5.1  & -71 & -103 & -345  & -229  & -76 & 26.3 & 31.3 & 26.9 & 31.8 & 50.2 & 58.8 & \textbf{\textcolor{LightCyan4}{0.74}} & \textbf{\textcolor{LightCyan4}{0.60}}  \\
    \end{tabular}
    \end{ruledtabular}
\end{table*}

\section{Results}
We define a few statistical error metrics that will be used to compare the performance of the density functionals analyzed (DFAs): the mean absolute error (MAE)
\begin{equation}
    \text{MAE} = \frac{1}{N} \sum_{i=1}^N |J_i^\text{DFA} - J_i^\text{exp}|,
\end{equation}
where $J_i^\text{DFA}$ is a magnetic coupling parameter computed with a DFA, and $J_i^\text{exp}$ is an experimentally determined value, serves as a general metric for accuracy. We assume $N$ quantities belong to a set.
The mean signed error (MSE) is
\begin{equation}
    \text{MSE} = \frac{1}{N} \sum_{i=1}^N (J_i^\text{DFA} - J_i^\text{exp}).
\end{equation}
When analyzed in conjunction with the MAE, the MSE is useful for determining the degree to which a DFA makes systematic errors.
The root-mean-square error (RMSE)
\begin{equation}
    \text{RMSE} = \left[\frac{1}{N} \sum_{i=1}^N (J_i^\text{DFA} - J_i^\text{exp})^2 \right]^{1/2},
\end{equation}
is a metric comparable to the MAE.
Finally, mean absolute relative error (MARE) is defined as
\begin{equation}
    \text{MARE} = \frac{1}{N} \sum_{i=1}^N |\frac{J_i^\text{DFA} - J_i^\text{exp}}{J_i^\text{exp}}|.
\end{equation}
Given the wide variation of experimental values for $J$ of different dimers, MARE prevents larger $J$ values from dominating the error metric at the cost of overemphasizing the smaller $J$, allowing small variations for the smaller $J$ to dominate.

Calculated values for $J$ of each combination of functional and molecule are provided in table \ref{tab:table2} along with experimental values.
MAE, MSE, RMSE, and MARE for $J$ of each functional are also provided in table \ref{tab:table2} and serve as metrics for the reliability of the tested XC functionals.
Error metrics are provided for both the set of five manganese dimers and the expanded set of all seven molecules.
For MAE, RMSE, and MARE, top performers are printed with bold text and ranked for smallest error with an \emph{Olympic medal} color scheme; italicized gold for 1st, silver for 2nd, and bronze for 3rd.
\begin{figure}[h]
\begin{tikzpicture}
\begin{axis}[
    ybar=0.5pt,
    bar width=6pt,
    ymax=50,
    xtick=data,
    ytick distance={5},
    ymin=0,
    symbolic x coords={SCAN, r$^2$SCAN, B3LYP, PBE0, TPSSh, lh07t-SVWN, lh07s-SVWN, lh12ct-ssir, lh12ct-ssif, lh14t-calPBE, lh20t},
       x tick label style={rotate=90, anchor=east},
    ylabel={MAE of calculated $J$ $($$cm^{-1}$$)$},
    nodes near coords,
    every node near coord/.append style = {rotate=90, anchor = west, font=\footnotesize},
        legend style={at={(0.5, -0.40)},
            anchor=north, legend columns=-1},
]
\usetikzlibrary{patterns}
\addplot [color=black, fill=black] coordinates {(SCAN, 12.1) (r$^2$SCAN, 7.1) (B3LYP, 19.3) (PBE0, 37.4) (TPSSh, 10.6)  (lh07t-SVWN, 20.0) (lh07s-SVWN, 15.6) (lh12ct-ssir, 25.3) (lh12ct-ssif, 31.3) (lh14t-calPBE, 19.9) (lh20t, 26.9)};
\addplot [color=black, pattern=north east lines] coordinates {(SCAN, 11.5) (r$^2$SCAN, 22.2) (B3LYP, 22.3) (PBE0, 43.7) (TPSSh, 14.5) (lh07t-SVWN, 19.0) (lh07s-SVWN, 15.9) (lh12ct-ssir, 23.5) (lh12ct-ssif, 31.2) (lh14t-calPBE, 16.1) (lh20t, 31.8)};
\legend{The 5 Mn$_2$ dimers$\:\:\:$, all 7 dimers}
\end{axis}
\end{tikzpicture}
\caption{\label{fig:MAE}Mean absolute error (MAE) for calculated $J$ $($$cm^{-1}$$)$ from functionals of the study with reference to experimental values.}
\end{figure}
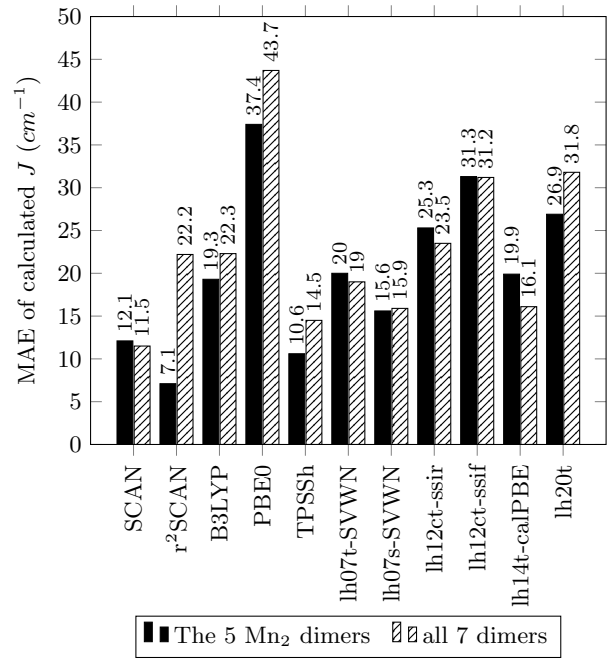
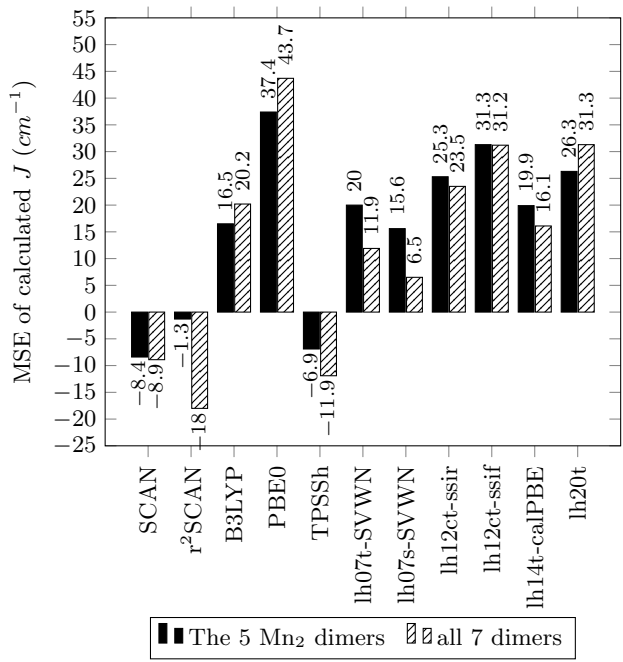
\begin{figure}[h]
\begin{tikzpicture}
\begin{axis}[
    ybar=0.5pt,
    bar width=6pt,
    ymin=-25,
    ymax=55,
    xtick=data,
    ytick distance={5},
    symbolic x coords={SCAN, r$^2$SCAN, B3LYP, PBE0, TPSSh, lh07t-SVWN, lh07s-SVWN, lh12ct-ssir, lh12ct-ssif, lh14t-calPBE, lh20t},
       x tick label style={rotate=90, anchor=east},
    ylabel={MSE of calculated $J$ $($$cm^{-1}$$)$},
    visualization depends on=y \as \myy,
    nodes near coords,
    nodes near coords style={rotate=90},
    every node near coord/.append style = {anchor ={90+sign(\myy)*90}, font = \footnotesize},
    nodes near coords style={xshift=0.06cm},
    legend style={at={(0.5, -0.40)},
            anchor=north, legend columns=-1},
]
\usetikzlibrary{patterns}
\addplot [color=black, fill=black] coordinates {(SCAN, -8.4) (r$^2$SCAN, -1.3) (B3LYP, 16.5) (PBE0, 37.4) (TPSSh, -6.9) (lh07t-SVWN, 20.0) (lh07s-SVWN, 15.6) (lh12ct-ssir, 25.3) (lh12ct-ssif, 31.3) (lh14t-calPBE, 19.9) (lh20t, 26.3)};
\addplot [color=black, pattern=north east lines] coordinates {(SCAN, -8.9) (r$^2$SCAN, -18.0) (B3LYP, 20.2) (PBE0, 43.7) (TPSSh, -11.9) (lh07t-SVWN, 11.9) (lh07s-SVWN, 6.5) (lh12ct-ssir, 23.5) (lh12ct-ssif, 31.2) (lh14t-calPBE, 16.1) (lh20t, 31.3)};
\legend{The 5 Mn$_2$ dimers$\:\:\:$, all 7 dimers}
\end{axis}
\end{tikzpicture}
\caption{\label{fig:MSE}Mean signed error (MSE) for calculated $J$ $($$cm^{-1}$$)$ from functionals of the study with reference to experimental values.}
\end{figure}
Bar graphs of MAE, MSE, RMSE, and MARE for a selection of DFAs are displayed in figures \ref{fig:MAE}, \ref{fig:MSE}, \ref{fig:RMSE}, and \ref{fig:MARE} respectively. 

In table \ref{tab:table2}, the BLYP and PBE GGAs and the TPSS meta-GGA have the highest MAE and unusably low accuracy, (fig. \ref{fig:MAE}).
MSE of $J$ for these three functionals (fig. \ref{fig:MSE}) shows a strong negative bias in accordance with expectations from previous studies \cite{velez2009density, PhillipsPeraltaMolSet}.
SIE along with the associated delocalization of electron densities is one of the most important causes of the $J$ errors of DFAs \cite{pederson2012self}.
BLYP, PBE, and TPSS all suffer significant SIE which drives delocalization of electronic density at the metal ions.
This corresponds to an underestimation of $U$ of equation \ref{J-contributions} and an expected anti-ferromagnetic bias.

In terms of accuracy as measured by MAE, the global hybrids B3LYP and PBE0 significantly improve over their GGA ancestors but still have considerable error.
For example, considering the manganese dimers, B3LYP's MAE ($19.3$ $cm^{-1}$) reduces BLYP's MAE ($107.0$ $cm^{-1}$) by a factor of $5.54$, and PBE0's MAE ($37.4$ $cm^{-1}$) reduces the MAE of PBE ($94.5$ $cm^{-1}$) by a factor of $2.53$. 
Despite this improvement, highly positive values for MSE indicate that the $J$ for B3LYP and PBE0 have been over-corrected.
This is likely because the the admixtures of SSD exact exchange in these two global hybrids are around $20$-$25\%$ --- too large for the $J$ of the considered systems, producing an over-localization of electron densities around the metal centers and overestimation of $U$.
TPSSh has a $10\%$ admixture of SSD exact exchange and works very well with an MAE of
$10.6$ $cm^{-1}$ that is roughly half of the MAE of B3LYP and an order of magnitude smaller than those of BLYP, PBE, and TPSS.  The success of TPSSh relative to the other global hybrids points to the previously mentioned problem regarding the empirical determination of the mixing parameter.

Though early local hybrids feature simple mixing functions with empirically-fitted parameters, the spatial-dependence of the mixing functions offer greater flexibility than the single admixture parameter of global hybrid functionals.
Unfortunately, the tested local hybrids are not superior to the global hybrids, especially when one considers the performance of TPSSh.
The most accurate local hybrids examined, lh07s-SVWN, lh07t-SVWN, and lh14t-calPBE, have MAE and RMSE slightly worse than TPSSh.
Considering MSE, all examined local hybrids show a positive ferromagnetic bias in $J$ indicating that the selective spatial dependence of the SSD admixture caused only marginal improvements upon the known bias of many global hybrids.

Interestingly, the SCAN family of meta-GGAs, here represented by SCAN and r$^2$SCAN, perform the best in terms of MAE, RMSE, and MARE over both the manganese set and expanded set of dimers.
This result challenges the idea that accurate calculation of magnetic coupling requires an admixture of SSD exact exchange \cite{OrioTPSSh, OrioHybrid, Pantazis, 07t, Jaramillo2003}.
MSE reveals a smaller bias than with most global or local hybrids. SCAN mis-characterizes the weakly ferromagnetic 1-TIPFAZ ($J$ = $10$ $cm^{-1}$) as antiferromagnetic with a $J^{SCAN}$ of $-4.0$ $cm^{-1}$, though the difference between $J^{exp.}$ and calculated $J^{SCAN}$ is one of the smallest.
In Pantazis's recent investigation that focused on double-hybrid performance on the same five manganese dimers \cite{Pantazis}, the accuracy of $J$ for SCAN was not as high as in our study, though for other functionals, our $J$ for each dimer were usually within a few $cm^{-1}$.
Given we used Turbomole \cite{Turbomole}, these differences could potentially be attributed to their use of ZORA relativistic scalar effects implemented in ORCA version 4 \cite{ORCA}, or the effects of grid settings.

\begin{figure}[t]
\begin{tikzpicture}
\begin{axis}[
    ybar=0.5pt,
    bar width=6pt,
    ymax=95,
    xtick=data,
    ytick distance={5},
    ymin=0,
    symbolic x coords={SCAN, r$^2$SCAN, B3LYP, PBE0, TPSSh, lh07t-SVWN, lh07s-SVWN, lh12ct-ssir, lh12ct-ssif, lh14t-calPBE, lh20t},
       x tick label style={rotate=90, anchor=east},
    ylabel={RMSE of calculated $J$ $($$cm^{-1}$$)$},
    nodes near coords,
    every node near coord/.append style = {rotate=90, anchor = west, font=\footnotesize},
        legend style={at={(0.5, -0.40)},
            anchor=north, legend columns=-1},
]
\usetikzlibrary{patterns},
\addplot [color=black, fill=black] coordinates {(SCAN, 8) (r$^2$SCAN, 20) (B3LYP, 35) (PBE0, 68.2) (TPSSh, 17) (lh07t-SVWN, 24) (lh07s-SVWN, 19) (lh12ct-ssir, 40) (lh12ct-ssif, 51) (lh14t-calPBE, 29) (lh20t, 50)};
\addplot [color=black, pattern=north east lines] coordinates {(SCAN, 12) (r$^2$SCAN, 50) (B3LYP, 42) (PBE0, 82.7) (TPSSh, 24) (lh07t-SVWN, 22) (lh07s-SVWN, 19) (lh12ct-ssir, 41) (lh12ct-ssif, 55) (lh14t-calPBE, 27) (lh20t, 59)};
\legend{The 5 Mn$_2$ dimers$\:\:\:$, all 7 dimers}
\end{axis}
\end{tikzpicture}
\caption{\label{fig:RMSE} The root-mean-square error (RMSE) of $J$ $($$cm^{-1}$$)$ versus experimental value for each functional of the study.}
\end{figure}
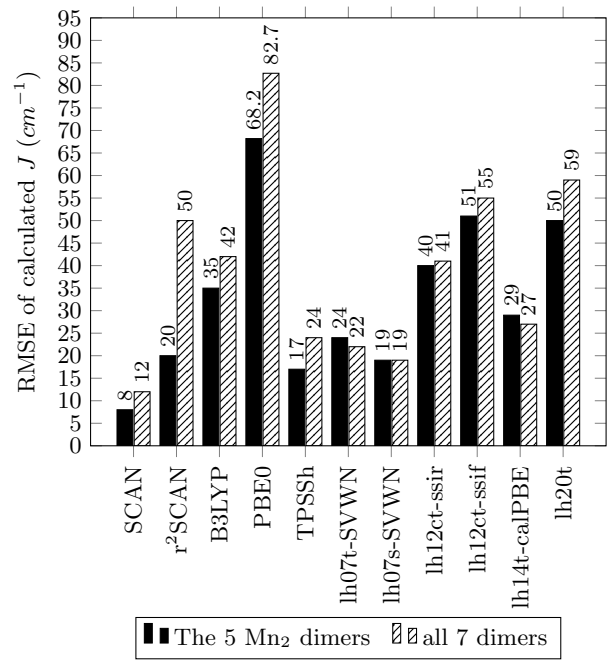
\begin{figure}[h]
\begin{tikzpicture}
\begin{axis}[
    ybar=0.5pt,
    bar width=6pt,
    ymax=1.5,
    xtick=data,
    ytick distance={0.1},
    ymin=0,
    symbolic x coords={SCAN, r$^2$SCAN, B3LYP, PBE0, TPSSh, lh07t-SVWN, lh07s-SVWN, lh12ct-ssir, lh12ct-ssif, lh14t-calPBE, lh20t},
       x tick label style={rotate=90, anchor=east},
    ylabel={MARE of calculated $J$},
    nodes near coords,
    every node near coord/.append style = {rotate=90, anchor = west, font=\footnotesize},
        legend style={at={(0.5, -0.40)},
            anchor=north, legend columns=-1},
]
\usetikzlibrary{patterns}
\addplot [color=black, fill=black] coordinates {(SCAN, 1.29) (r$^2$SCAN, 0.42) (B3LYP, 0.85) (PBE0, 1.15) (TPSSh, 1.03) (lh07t-SVWN, 1.33) (lh07s-SVWN, 1.09) (lh12ct-ssir, 0.85) (lh12ct-ssif, 1.11) (lh14t-calPBE, 0.97) (lh20t, 0.74)};
\addplot [color=black, pattern=north east lines] coordinates {(SCAN, 0.94) (r$^2$SCAN, 0.38) (B3LYP, 0.64) (PBE0, 0.90) (TPSSh, 0.77) (lh07t-SVWN, 0.98) (lh07s-SVWN, 0.80) (lh12ct-ssir, 0.64) (lh12ct-ssif, 0.84) (lh14t-calPBE, 0.71) (lh20t, 0.60)};
\legend{The 5 Mn$_2$ dimers$\:\:\:$, all 7 dimers}
\end{axis}
\end{tikzpicture}
\caption{\label{fig:MARE}Mean absolute relative error (MARE) for calculated $J$ from functionals of the study with reference to experimental values.}
\end{figure}
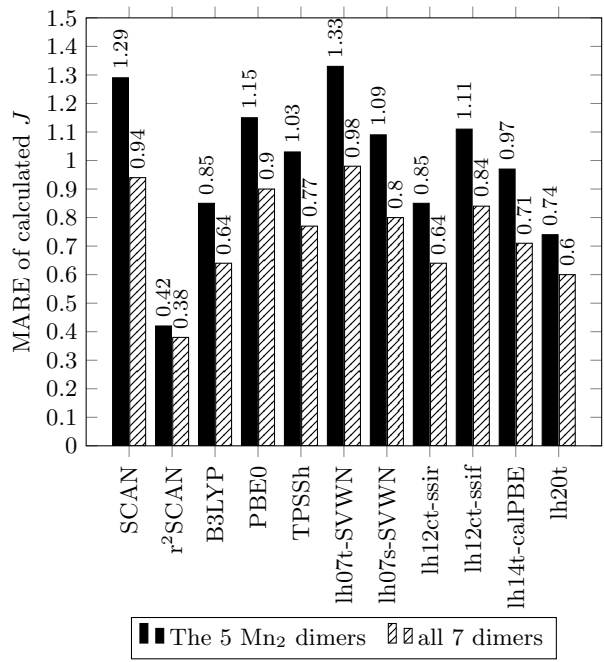

\begin{figure}[h]
\includegraphics[width=0.48\textwidth, height=0.38\textwidth]{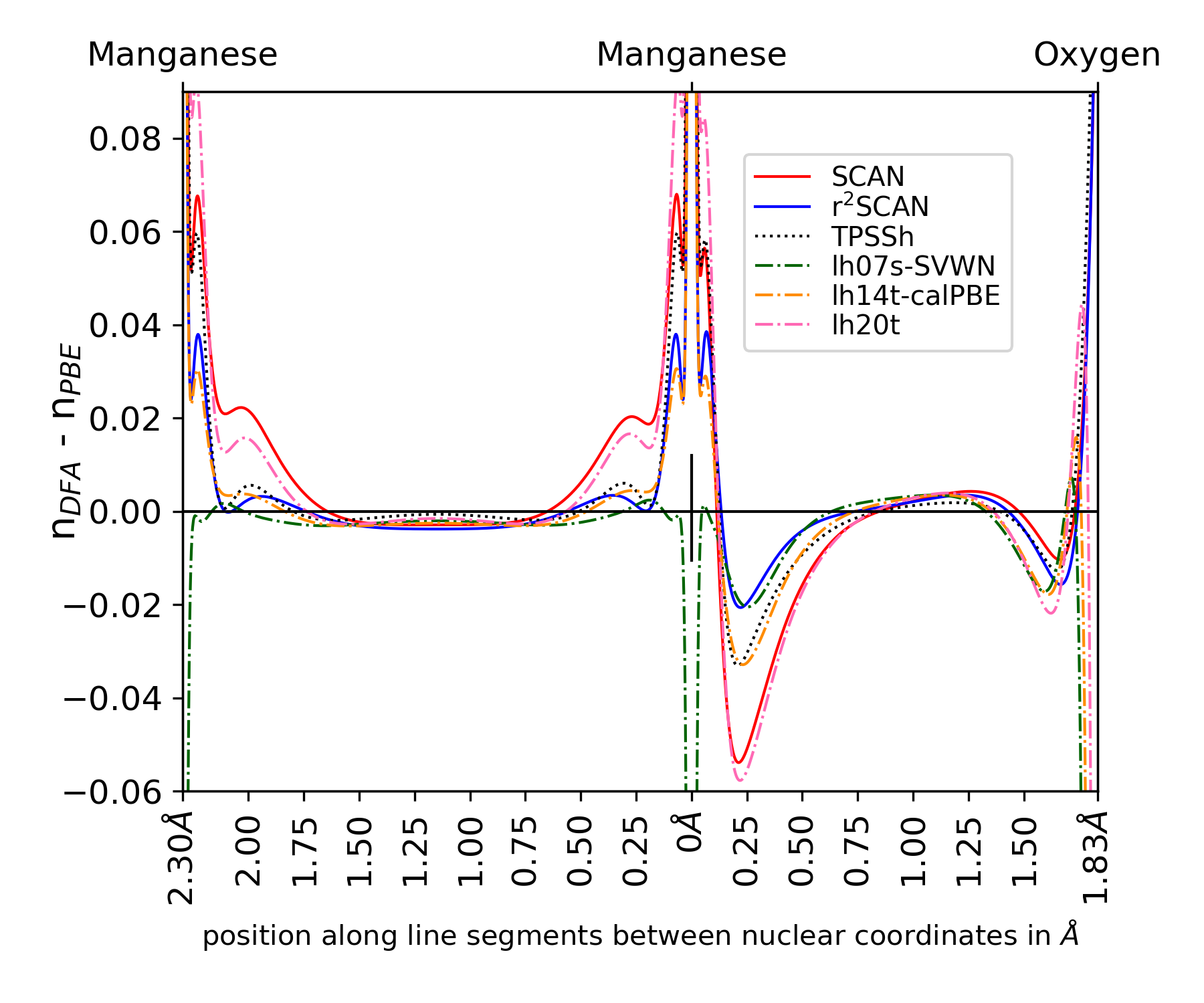}
\caption{\label{fig:5VADDAF_td}The total density from PBE was subtracted from the functionals of the study and plotted along inter-nuclear line segments with distance given in \AA. The high spin state of 5-VADDAF is shown.}
\end{figure}

SCAN showed the lowest RMSE for both sets (figure \ref{fig:RMSE}) indicating that it has both high accuracy and a predictable bias in $J$.
TPSSh and the local hybrids lh07t-SVWN and lh07s-SVWN also show low RMSE error for both sets.

The MARE metric is intended to de-emphasize the dimers with larger $J$ values, though the two dimers with the two smallest experimental $J$, 1-TIPFAZ ($J_{exp} = 10$ $cm^{-1}$) and 2-GEFKAD ($J_{exp} = -3.4$ $cm^{-1}$) are arguably over-emphasized, as the error in the calculated $J$ does not scale uniformly with the magnitude of experimental $J$ (figure \ref{fig:MARE}). Regardless of this shift in bias, the change of emphasis to dimers with smaller experimental $J$ does not affect the accuracy of r$^2$SCAN, which has the best error performance for both the manganese set and the expanded set. The newest local hybrid tested, lh20t, performs well, producing the second best MARE for both tests sets. SCAN drifts back in performance when considering MARE compared to the other measures of error, largely due to the performance on 2-GEFKAD. This result is in line with previous observations that SCAN performs very well with dimers having larger $J$, and not as well with dimers having smaller $J$ \cite{Pantazis}.

To help visualize electron delocalization and the effects of SIE, we plot the density difference between DFAs and PBE, letting PBE serve as a baseline for comparison of density distributions, see figure \ref{fig:5VADDAF_td}.
Several successful functionals with superior error statistics are chosen for the density difference plot.
We use the high spin state of 5-VADDAF as a representative case for this analysis.
The density differences plotted here are along the line connecting the two metal nuclei, and the line connecting a metal nucleus to the bridging oxygen nucleus.

Figure \ref{fig:5VADDAF_td} shows that all the considered functionals except for lh07s-SVWN increase the electron density at the manganese core region (from $0$ to roughly $0.10$\AA$ $ $ $from the manganese nuclear positions) in comparison with PBE, indicating more localized electron density. 
In the inter-manganese region on the left side of figure \ref{fig:5VADDAF_td}, most of the functionals show an increased localization of density to a non-bonding valence-orbital region of the manganese ions, a result in line with the expected large unphysical delocalization of these unpaired, non-ionized manganese electrons given by PBE.
Along the path between manganese and oxygen atoms where ionic bonding occurs, all functionals tend to deplete total density in the manganese valence region.
These results give a rough qualitative insight into how improvements over PBE are generally \emph{re}locating density to achieve their successes in correcting delocalization of unpaired electrons and the associated error in the magnetic contributions to the energy.
However, quantitative connections between the MAE and the degree of localization for different functionals are hard to analyze.
Similar plots for spin-density and a spin-resolved electron localization function justify identification of unpaired electron density and can be found in the supporting information.

\section{Conclusion} 
Previous studies have shown that GGA functionals and some meta-GGA functionals over-stabilize the anti-ferromagnetic state, while hybrid functionals can improve upon GGAs, but they can over-correct, producing a ferromagnetic bias \cite{Pantazis, 07t, bandeira2012calculation, Jaramillo2003}.
This translates in exchange couplings that are too antiferromagnetic for GGAs and meta-GGAs, and slightly too ferromagnetic for hybrid functionals.
The results for TPSSh and the SCAN family functionals are exceptions to this pattern.
We see that two generalities are challenged; the SCAN and r$^2$SCAN functionals, without mixing SSD exact exchange, show a high level of accuracy and reliability in the determination of $J$ couplings while TPSSh doesn't have a ferromagnetic bias, producing results with a similar accuracy and bias to the SCAN family of meta-GGAs.
As a practical point, for the calculation of J couplings in high-nuclearity complexes, SCAN and r$^2$SCAN offer a computationally attractive option since they do not require the evaluation of Hartree-Fock type exchange.
Since the number of considered molecular complexes is rather small, these observations have to be tested for a larger and more inclusive set of magnetic molecular complexes, but it should serve as an indication of the general trends expected.

In the search for candidate quantum materials and their potential application, the large size of SMMs and the potential use of high throughput methods places an emphasis on efficiency with reasonable accuracy.
The use of meta-GGAs SCAN and r$^2$SCAN for MCTMs may produce the highest available accuracy for DFT while not using resources on arguably unnecessary SSD exact exchange calculations.

\section*{Acknowledgements}
H.C.F., J.W.F., and J.S. acknowledge the support of the U.S. DOE, Office of Science, Basic Energy Sciences Grant No. DESC0019350 (core research).
M.R.P. was supported by DOE-BES M2QM EFRC under award number DE-SC00193.
J.E.P. acknowledges support from the Office of Basic Energy Sciences, U.S. Department of Energy (Grant DE-SC0005027).
We thank Dimitrios Pantazis for his structure files.

\FloatBarrier
\twocolumngrid
\bibliography{main.bib}
\end{document}